\documentstyle[12pt]{article}

\renewcommand{\eG}{e^{ {\vec\xi}.{\vec G}}}
\renewcommand{\no}{{\bf :}}
\begin{document}

\begin{titlepage}

\vskip .6in

\begin{center}
{\Large {\bf   The Berkovits-Vafa Construction and   }}\\[10pt]
{\Large {\bf  Nonlinear Realizations}}
\end{center}

\normalsize
\vskip .6in

\begin{center}

I. N. McArthur
\par \vskip .1in \noindent
 {\it Department of Physics, The University of Western Australia}\\
{\it Nedlands, Australia.  6009}

\end{center}
\vskip 3cm

\begin{center}
{\large {\bf ABSTRACT}}\\
\end{center}

Berkovits and Vafa recently showed that  critical string theories can be
considered as  critical superstring theories with a special choice of
background. The embedding of the Virasoro algebra into the super-Virasoro
algebra  involved in this construction has been extended to the noncritical
case by Berkovits and Ohta. It is shown that the resulting nonlinear
realization of the super-Virasoro algebra can be interpreted using standard
techniques from the theory of nonlinear realizations. This extends earlier
work of Kinitomo.
\vspace{9cm}
\noindent

\end{titlepage}

\noindent
{\bf 1. Introduction}

Recently, Berkovits and Vafa showed that critical strings can be considered
as special choices of vacuua for critical N=1 superstring theories \cite{BV}.
 The essence of the construction is an embedding of the c=26 Virasoro algebra
of the critical string into an N=1 superconformal algebra which is nonlinearly
realized in terms of the Virasoro generators and a set of fermionic ghosts
$(b,c)$ of conformal weight $(\frac32, -\frac12).$ This construction has
since been generalized to higher N and to noncritical dimensions \cite{BO}.

Here, it is shown that the embedding of the noncritical Virasoro algebra into
a nonlinearly realized N=1 superconformal algebra can be obtained using the
 standard theory of nonlinear realizations, as applied to the super-Virasoro
algebra. This should also help to clarify an iterative procedure outlined by
Berkovits and Ohta \cite{BO} for carrying out this embedding.

In \cite{Kun},  the
theory of nonlinear realizations is also applied to this problem. However,
by using an approach based upon superconformal fields in superspace,
the author
is effectively restricting to the case of a super-Virasoro algebra with
vanishing central charge, with the result that much of the additional
structure obtained
using the approach detailed below has to be added in by hand in the form
of ``quantum corrections''.

We begin with a quick review of the elements of the theory of nonlinear
realizations
of ordinary groups, following the exposition of Coleman, Wess and Zumino
\cite{CWZ}; see also \cite{WB}. If $H$ is a subgroup of a Lie group $G$, and
if $({\vec T}, {\vec X})$ denote the generators of the Lie algebra of $G$ with
${\vec T}$ the generators of the Lie algebra of $H,$ then an arbitrary
element $g\in G$ can be decomposed in the form $g = e^{{\vec \xi}.{\vec X}}
e^{{\vec u}.{\vec T}}.$ In particular, $e^{{\vec \xi}.{\vec X}}$
parameterizes a slice through $G$ in a neighbourhood of the identity
which is isomorphic to $G/H.$ A nonlinear action of the
group on the manifold $G/H$  is determined by
\begin{equation}
g \,e^{{\vec \xi}.{\vec X}} =  e^{{\vec \xi}\, '.{\vec X}}
e^{{\vec u}\,'.{\vec T}},
\end{equation}
where ${\vec \xi}'$and ${\vec u}\,'$ depend on ${\vec \xi}$ and $g \in G.$
Further, if $\psi \rightarrow D(h) \psi$ for $h\in H$ is a (linear)
representation
of $H$ on some space $V$ with elements $\psi$, then the transformation
$(\xi, \psi) \rightarrow
(\xi', D(e^{{\vec u}\,'.{\vec T}})\psi)$ also determines a nonlinear
realization of the group, this time on a space which is locally the tensor
product
of $G/H$ and $V.$ Note that while the transformation $\xi \rightarrow \xi'$
generates a nonlinear realization in its own right, this is not true of
$\psi \rightarrow D(e^{{\vec u}\,'.{\vec T}})\psi,$ because ${\vec u}\,'$
depends
on $\xi.$

We now extend this construction to an infinite dimensional setting by
 replacing the
Lie algebra of $G$ with the super-Virasoro algebra of central charge $d$
and considering  its Virasoro subalgebra\footnote{The Wakimoto
free field realization of Kac-Moody algebras with arbitrary central charge
was interpreted in a related manner in \S 5 of \cite{Felder}.}. In the
Ramond sector, the super-Virasoro algebra has even generators $L_n$
and odd generators $G_n$ which satisfy the relations
\begin{eqnarray}
\,[L_n, L_m] &=& (n-m) L_{n+m} + \frac{d}{12} (n^2 - n) \delta_{n+m,0}{\bf 1}
\nonumber \\
\, [L_n, G_m] &=& (\frac12 n - m)G_{n+m} \nonumber \\
\, \{G_n, G_m\} &=& 2L_{n+m} + \frac{d}{3} (n^2 - \frac14) \delta_{n+m,0}{\bf
1}
\label{sVir}
\end{eqnarray}
and which form the modes of the stress tensor and supercurrent,
$T(z) = \sum_n L_n z^{-n-2},$ $G(z) = \sum_n G_n z^{-n-3/2}.$
The generators $L_n$ and the central element ${\bf 1}$ are the analogues of
 ${\vec T}$ in the earlier construction, and the generators
$G_n$ are the analogues of  ${\vec X}.$ We are thus led to a
consideration of the action of group elements corresponding to the above
algebra on $\exp(\sum_m \xi_m G_m) \equiv e^{ {\vec \xi}.{\vec G}},$
$$g\, \eG = \exp(\sum_m \xi_m' G_m) \, \exp(\sum_m u_m' L_m + u'{\bf1}).$$
 The $\xi_m$ are Grassmann
parameters as $G_m$ are odd elements of the algebra. Actually, to obtain
generators which are realized by a right action, we consider
$$\eG \, g= \exp(\sum_m u_m' L_m + u'{\bf1}) \, \exp(\sum_m \xi_m' G_m) \,.$$
In infinitesimal form, this corresponds to expressing $\eG\,L_n$ and
$\eG\,G_n$ as linear combinations of terms of the form $L_k\,\eG $
and ${\bf 1}\,\eG,$ and terms involving derivatives with respect to $\xi_k$
(the latter generating transormations $\xi \rightarrow \xi'$).

In the case of $\eG\, L_n,$ it can be written $[\eG, \,L_n ] + L_n\, \eG .$
Due to the fact
that the generators $G_m$ do not anticommute among themselves, the commutator
is equivalent to
\begin{eqnarray*}
& &  \int_0^1 ds\, e^{s{\vec \xi}.{\vec G}} \, [ \sum_m \xi_m G_m,\, L_n]
\,e^{(1-s){\vec \xi}.{\vec G}}  \\
&=& -\sum_m (\frac12n-m) \,\xi_m \int_0^1 ds\, e^{s{\vec \xi}.{\vec G}} \,
 G_{m+n}\,e^{(1-s){\vec \xi}.{\vec G}}  \\
&=& -\sum_m (\frac12n-m)\, \xi_m \frac{\partial}{\partial \xi_{m+n}} \, \eG.
\end{eqnarray*}
This yields an expression of the required form:
\begin{equation}
\eG\, L_n = L_n\, \eG - \sum_m (\frac12n-m) \, \xi_m
\frac{\partial}{\partial \xi_{m+n}} \, \eG.
\label{Ln}
\end{equation}

The case $\eG \, G_n$ is more difficult. It is actually simpler to begin with
\begin{eqnarray*}
\frac{\partial}{\partial \xi_{n}} \, \eG &=&
\int_0^1 ds\, e^{(1-s){\vec \xi}.{\vec G}} \, G_n\,
e^{s{\vec \xi}.{\vec G}}\\
&=& \int_0^1 ds\, e^{(1-s){\vec \xi}.{\vec G}}\, [G_n,\,
e^{s{\vec \xi}.{\vec G}}] + \eG \, G_n.
\end{eqnarray*}
By evaluation of the commutator, the first term is equivalent to
$$ -2\sum_m \xi_m \int_0^1 ds\,s \int_0^1 dt\, e^{(1-st){\vec \xi}.{\vec G}} \,
 L_{m+n}\,e^{st{\vec \xi}.{\vec G}} -
\frac{d}{6} (n^2 - \frac14) \xi_{-n} {\bf 1}\,\eG.$$
The generator $L_{m+n}$ can be moved to the left with the introduction of a
new commutator, which can be evaluated as before to give
\begin{eqnarray*}
& &\sum_{m,p} (m+n-2p)\, \xi_m \xi_p \int_0^1 ds\,s \int_0^1 dt\,
 \biggl( \frac{\partial}{\partial \xi_{m+n+p}}e^{(1-st){\vec \xi}.{\vec G}}
\biggr)\,
 \,e^{st{\vec \xi}.{\vec G}}\\
&=& \frac12 \sum_{m,p} (m+n-2p) \,\xi_m \xi_p
 \frac{\partial}{\partial \xi_{m+n+p}} \eG \\
& & \, -\sum_{m,p} (m+n-2p)\, \xi_m \xi_p \int_0^1 ds\,s^2 \int_0^1 dt\, t
\, e^{(1-st){\vec \xi}.{\vec G}}\int_0^1 du\, e^{(1-u)st{\vec \xi}.{\vec G}} \,
 G_{m+n+p}\,e^{ust{\vec \xi}.{\vec G}}.
\end{eqnarray*}
Again, in the last term, the generator $G_{m+n+p}$ can be moved to the right
with the introduction of a commutator, which is of a form which has already
been evaluated. One of the resulting terms is of the form
$$\sum_{m,p,q} (m+n-2p)\, \xi_m \xi_p \xi_q \, {\hat O}_1 L_{m+n+p+q} {\hat
O}_2,$$
with ${\hat O}_1$ and ${\hat O}_2$ operators. Using $\xi_m \xi_p =
 - \xi_p \xi_m,$ this is equivalent to $$\frac32 \sum_{m,p,q} (m-p)\,
\xi_m \xi_p \xi_q \,{\hat O}_1 L_{m+n+p+q} {\hat O}_2,$$ which can be seen
to vanish
by interchanging the dummy variables $m, p, q$ cyclically and using the
symmetry properties of $\xi_m \xi_p \xi_q.$

Putting all the above together, one finds:
\begin{eqnarray*}
\eG\, G_n &=& \frac{\partial}{\partial \xi_n} \eG + \sum_m \xi_m  L_{m+n} \,\eG
\nonumber \\
& & \,\
- \frac34 \sum_{m,p} (m-p) \,\xi_m \xi_p \frac{\partial}
{\partial \xi_{m+n+p}}\eG +\frac{d}{6}
(n^2 - \frac14) \, \xi_{-n}\, {\bf 1} \eG  \nonumber \\
& & \, \, - \frac{d}{48} \sum_{m,p} (m-p) \biggl( (m+n+p)^2 - \frac14 \biggr)
\, \xi_m \xi_p \xi_{-(m+n+p)}\, {\bf 1}\eG \nonumber \\
& & \, \,  +\frac14 \sum_{m,p}(m-p) \,\xi_m \xi_p \,
 \eG\, G_{m+n+p}.
\end{eqnarray*}

The last term involves $ \eG \, G_{m+n+p},$ which is of the same form as
the left hand side of the equation,
so  this expression can be
evaluated iteratively. However, the process terminates after one iteration,
and leads simply to a correction of some of the coefficients involved.
In particular,
$$\sum_{m,p,q}(m-p)\, \xi_m \xi_p \xi_q\,  L_{m+n+p+q} \eG $$ vanishes
by an earlier argument. The terms in the iterated expression
involving four or more $\xi$ factors are of the form $$ \sum_{m,p,q}(m-p) \,
(q-r)\,\xi_m \xi_p \xi_q \xi_r\, \eG f(m+n+p+q+r).$$ By cycling the dummy
variables $m,p$ and $q,$ this expression is seen to vanish. So the final
result is:
\begin{eqnarray}
\eG\, G_n &=& \frac{\partial}{\partial \xi_n} \eG +  \sum_m \xi_m L_{m+n}\, \eG
- \frac12 \sum_{m,p} (m-p) \, \xi_m \xi_p
\frac{\partial}{\partial \xi_{m+n+p}}\eG
\nonumber \\
& & \, \, +\, \frac{d}{48} \sum_{m,p} (m-p) \,\biggl( (m+n+p)^2 - \frac14
\biggr)
\, \xi_m \xi_p \xi_{-(m+n+p)}\,{\bf 1} \eG \nonumber \\
& & \,\,+\,  \frac{d}{6} (n^2 - \frac14)\, \xi_{-n} \,{\bf 1} \eG.
\label{Gn}
\end{eqnarray}

Applying the standard theory of nonlinear realizations to the results
(\ref{Ln})
and (\ref{Gn}), it
would follow that if we have a representation of the generators of the Virasoro
algebra on a space of states $\psi,$ then a nonlinear realization of
the super-Virasoro algebra on the space $({\vec \xi}, \psi)$ is generated
by
\begin{eqnarray}
{\tilde L_n} &=& L_n - \sum_m (\frac12 n - m)\, \xi_m \, \frac{\partial}
{\partial \xi_{m+n}}\nonumber \\
{\tilde G_n} &=& \frac{\partial}{\partial \xi_n} + \sum_m \xi_m L_{m+n}
- \frac12 \sum_{m,p} (m-p) \,\xi_m \xi_p \frac{\partial}
{\partial \xi_{m+n+p}}\nonumber \\
& &
\, \, + \frac{d}{48} \sum_{m,p} (m-p) \,\biggl( (m+n+p)^2 - \frac14 \biggr)
\, \xi_m \xi_p \xi_{-(m+n+p)}\nonumber \\
& & \, \, + \frac{d}{6} (n^2 - \frac14)\, \xi_{-n},
\label{nonlin}
\end{eqnarray}
where we have used $L_n$ to denote the representation of the Virasoro
generators on the space of states $\psi.$ Identifying $\xi_{-m}$ and
$\frac{\partial}{\partial \xi_m}$ with fermionic mode operators $c_m$ and
$b_m$ respectively,
then $\{ \frac{\partial}{\partial \xi_m}, \xi_{-n} \} = \delta_{m+n,0}$ is
equivalent to the standard ghost anticommutation relations $\{ b_m,
c_n \} = \delta_{m+n,0}.$ With $c(z) = \sum_n c_n z^{-n+\frac12}$ and
 $b(z) = \sum_n b_n z^{-n-\frac32},$ the generators ${\tilde L_n}$ and
${\tilde G_n}$ can be identified as mode operators for the fields
\begin{eqnarray}
{\tilde T}(z) &=& \sum_n {\tilde L_n} z^{-n-2} \nonumber \\
&=& T(z) + \frac32 \partial c(z) b(z) + \frac12 c(z) \partial b(z)\nonumber \\
{\tilde G}(z) &=& \sum_n {\tilde G_n} z^{-n-\frac32} \nonumber \\
&=& b(z) + c(z) T(z) + c(z) \partial c(z) b(z) \nonumber \\
& &\, \, \,-\frac{d}{24} c(z) \partial c(z)
\partial^2 c(z) + \frac{d}{6} \partial^2 c(z).
\label{nonlin1}
\end{eqnarray}
The assignment of conformal weight $-\frac12$ to $c(z)$ is natural as
${\vec \xi}.{\vec G} = \oint dz \, c(z) G(z)$ is then a conformally invariant
combination.

Although (\ref{nonlin})  provides a nonlinear realization
of the super-Virasoro algebra with central charge $d$, the operators
(\ref{nonlin1}) are not of the same form as those for the noncritical
embedding in \cite{BO}.
  This is because the expressions
$\partial c(z) b(z),$ $c(z) \partial b(z)$ and $c(z) \partial c(z) b(z)$ in
(\ref{nonlin1}) are not normal ordered, and the normal ordered generators
require ``quantum correction'' to produce a closed algebra. In fact, the terms
requiring normal ordering are those which produce the transformation $\xi
\rightarrow \xi',$ and so the completion of the Berkovits-Ohta noncritical
embedding  is reduced to the task of promoting the nonlinear
realization on the ghost Fock space from a ``classical level'' with
generators
\begin{eqnarray*}
{\tilde T}(z) &=& \frac32   \partial c (z) b(z)
 + \frac12  c(z) \partial b(z) \\
{\tilde G}(z) &=& b(z) +   c(z) \partial c(z) b(z)
\end{eqnarray*}
to a ``quantum level'' involving normal ordered operators. As is evident from
the original critical embedding of Berkovits and Vafa, this is indeed possible,
with
\begin{eqnarray*}
{\tilde T}(z) &=& -\frac32 \no b \, \partial c \no (z)
 - \frac12 \no \partial b\,  c \no (z)
 + \frac12 \partial^2(c \, \partial c)(z) \\
{\tilde G}(z) &=& b(z) + \no b \, \no c \,\partial c\no \no (z) +
\frac{26}{24} c\, \partial c \,
\partial^2 c (z)  - \frac{11}{6} \partial^2c(z)
\end{eqnarray*}
generating a super-Virasoro algebra with central charge $-11.$ It should be
noted that the field content  of the corrections to ${\tilde G}$ is of the
same form as those due to the nonvanishing central charge in the Virasoro
sector, and those in ${\tilde T}$ follow from the
operator product expansion ${\tilde G}(z) \, \, \,
{\tilde G}(w).$ It is then a matter of choosing the coefficients to ensure
that the generators satisfy the super-Virasoro algebra.

 Substituting these
expressions for the generators of the transformations on the ghost
Fock space in (\ref{nonlin1}) yields
\begin{eqnarray*}
{\tilde T}(z) &=& T(z) -\frac32 \no b \, \partial c \no (z)
 - \frac12 \no \partial b\, c \no (z)
 + \frac12 \partial^2(c\, \partial c)(z) \\
{\tilde G}(z) &=& b(z) + cT(z) + \no b\, \no c\,\partial c\no \no (z) -
\frac{(d-26)}{24} c\, \partial c \,
\partial^2 c (z)  + \frac{(d-11)}{6} \partial^2 c(z).
\end{eqnarray*}
This is the result of Berkovits and Ohta for the noncritical embedding of the
conformal algebra of central charge $d$ into the N=1 superconformal algebra
of central charge $(d-11).$

In summary, we have shown that the Berkovits-Ohta noncritical embedding of
the Virasoro algebra into the N=1 superconformal algebra is related to
standard techniques for producing nonlinear realizations. In particular, this
extends the results of \cite{Kun}, in that it allows a systematic construction
of {\it all} the terms  which act on the
space on which the original Virasoro algebra is realized linearly, although
corrections due to normal ordering in the ghost sector still have to be
computed by other means. It also casts some light on the iterative construction
in \cite{BO}, where a similarity transformation is used
  to remove dependence on an auxiliary stress tensor and
supercurrent which are introduced. The generator of the similarity
transformation
is  the quantity $e^{\oint cG}$ which appears in the construction of the
nonlinear
realization.

\end{document}